\documentclass[12pt,preprint]{aastex}

\usepackage{natbib,emulateapj5}

\def\ltsima{$\; \buildrel < \over \sim \;$}
\def\simlt{\lower.5ex\hbox{\ltsima}} 
\def\gtsima{$\; \buildrel > \over \sim \;$}
\def\simgt{\lower.5ex\hbox{\gtsima}} 

\def\feka{Fe K$\alpha$}
 
\def\xmm{{\it XMM-Newton}} 
\def\asca{{\it ASCA}} 
 
\def\rxte{{\it RXTE}} 
\def\sax{{\it BeppoSAX}}

\def\lum{{\rm erg~s^{-1}}}

\begin{document}

\def\aj{\rm{AJ}}                   
\def\araa{\rm{ARA\&A}}             
\def\apj{\rm {ApJ}}                
\def\apjl{\rm{ApJ}}                
\def\apjs{\rm{ApJS}}               
\def\apss{\rm{Ap\&SS}}             
\def\aap{\rm{A\&A}}                
\def\aapr{\rm{A\&A~Rev.}}          
\def\aaps{\rm{A\&AS}}              
\def\mnras{\rm{MNRAS}}             
\def\nat{\rm{Nature}}              
\def\pasj{\rm{PASJ}}    	   
\def\procspie{\rm{Proc.~SPIE}}     

\title{X-Ray Spectra of Intermediate-Luminosity, Radio-Loud Quasars } 

\author{Christa A. Hasenkopf, Rita M. Sambruna\altaffilmark{1}, 
and Michael Eracleous}

\affil{Department of Astronomy \& Astrophysics, The Pennsylvania State
University, 525 Davey Laboratory, University Park, PA 16802}

\email{cah@astro.psu.edu, rms@physics.gmu.edu, mce@astro.psu.edu}

\altaffiltext{1}{Current address: Department of Physics \& Astronomy
and School of Computational Science, George Mason University,
Fairfax, VA 22030}

\begin{abstract}

We present new hard X-ray spectra of three radio-loud AGNs of
moderately high X-ray luminosity ($L_{\rm X}\approx 10^{45}~\lum$;
PKS~2349--01, 3C~323.1, and 4C~74.26) obtained with \asca\ and
\sax. The X-ray continua are described in all three cases with a power
law model with photon indices $\Gamma\approx1.85$, modified at low
energies by absorption in excess of the Galactic, which appears to be
due to neutral gas. At higher energies, an \feka\ emission line is
detected in PKS~2349--01 and 4C~74.26, and is tentatively detected in
3C~323.1. The equivalent widths of the lines are consistent, albeit
within large uncertainties, with the values for radio-quiet AGN of
comparable X-ray luminosity. The \feka\ line is unresolved in
4C~74.26. In the case of PKS~2349--01, however, the inferred
properties of the line depend on the model adopted for the continuum:
if a simple power-law model is used, the line is resolved at more than
99\% confidence with a full width at half maximum corresponding to
approximately 50,000~km~s$^{-1}$ and a rest-frame equivalent width of
$230\pm 120$~eV, but if a Compton ``reflection'' model is used the
line is found to be a factor of 2 weaker, for an assumed full width at
half maximum of 50,000~km~s$^{-1}$. In 4C~74.26, a strong Compton
``reflection'' component is detected. Its strength suggests that the
scattering medium subtends a solid angle of $2\pi$ to the illuminating
source. Overall, the spectral indices of these radio-loud quasars are
remarkably similar to those of their radio-quiet counterparts.  On the
other hand, if the absorber is indeed neutral, as our results suggest,
this would be consistent with the typical properties of radio-loud
AGNs.

\end{abstract}


\section{Introduction}
\noindent A major outstanding question in our ongoing efforts to
understand the central engines of active galactic nuclei (AGNs) is the
origin of the dichotomy between radio-loud and radio-quiet objects.
Although both radio-loud and radio-quiet AGNs are thought to be
powered by accretion onto a supermassive black hole, it is not known
why AGNs in the former class can produce collimated, relativistic
radio jets, which power the giant radio lobes at large distances from
their host galaxies, while AGNs in the latter class cannot
\citep[see][for a review]{l97}. One possible explanation is that the
central engines of the two classes of objects are the same, but
systematic differences in the properties of the host galaxies are the
cause of the observed dichotomy \citep[e.g.,][]{bl95,fr95}.  Another
possibility is that there is a systematic difference between the
central engines of radio-loud and radio-quiet AGNs, which is
intimately related to their ability to form jets. Such a difference
could be the size of the accretion disk \citep[invoked in some wind
models; e.g.,][]{bp82}, the spin of the black hole \citep[thought to
be the ultimate power source of the jets; e.g.,][]{bz77,wc95,m99}, or
the structure of the inner accretion disk \citep[an ion torus or
advection-dominated accretion flow, a.k.a.  ADAF, which allows some of
the accreting matter to form an outflow; e.g.,][]{r82,ny95,bb99}.

X-ray observations are a means by which one can investigate the last
two of the above possibilities, since the X-rays are produced in the
inner regions of the accretion flow. In particular, one can look for
systematic differences in the X-ray spectral properties of radio-loud
and radio-quiet AGNs that may indicate differences in the structure of
the inner accretion disk or the spin of the central black hole. Thus
we and other groups have been carrying out systematic studies of the
X-ray properties of radio-loud AGNs using archival as well as
proprietary \asca\ and \rxte\ observations. These studies have
confirmed previously suspected systematic differences between
radio-loud and radio-quiet AGNs \citep*{z95,w98,sem99,esm00}. In
particular, two spectral features that are the hallmarks of the X-ray
spectra of Seyfert galaxies \citep[e.g.,][]{np94,n97b}: the
fluorescent Fe K$\alpha$ line at 6.4~keV and the Compton ``reflection''
hump at energies greater than 10~keV, are considerably weaker in the
spectra of broad-line radio galaxies \citep[hereafter
BLRGs;][]{z95,w98,sem99,esm00}. The above observational results
suggest that the structure of the inner accretion disk of BLRGs is
different from that of Seyfert galaxies, although the nature of this
difference remains to be understood in detail. Another potentially
important trend is the X-ray Baldwin effect \citep{n97c}, according to
which the Fe K$\alpha$ lines of luminous quasars are systematically
weaker (i.e., of lower equivalent width; hereafter EW) than those of
the lower-luminosity Seyfert galaxies.  Because the sample of objects
studied by \citet{n97c} includes a mix of radio-loud and radio-quiet
quasars (hereafter QSRs and QSOs, respectively), it is not yet clear
whether and how such an effect applies to radio-loud and radio-quiet
AGNs separately. Finally, the immediate environment of the X-ray
source appears to be different in radio-loud and radio quiet AGNs as
evidenced by the properties of absorbing material along the line of
sight. More specifically, the signature of a ``warm'' (i.e., ionized)
absorber, seems to be ubiquitous in Seyfert galaxies
\citep[e.g.,][]{r97,g98} but rare in broad-line radio galaxies
\citep{sem99}. Instead, radio-loud AGNs seem to display the signature
of significant absorption by neutral matter in their X-ray spectra.

Here we present the results of new observations of three
intermediate-luminosity QSRs (3C~323.1, PKS~2349--01, 4C~74.24;
$L_{\rm X}\sim 10^{45}~{\rm erg~s^{-1}}$) with \asca\ and \sax, which
were carried out as part of our larger program. The targets were
specifically selected to bridge the gap in luminosity between the
nearby and well-studied BLRGs and the more luminous QSRs included in
the sample of \citet{n97c}. All three are associated with large,
double-lobed radio sources \citep{a85,k94,r88}, which ensures that our
line of sight is not close to the axis of their radio jet, thus the
X-rays we observe are not contaminated by beamed emission from the jet
itself. The goal of the observations was to investigate the properties
of intermediate-luminosity QSRs and to check whether the systematic
differences between Seyfert galaxies and BLRGs extend to higher
luminosity objects. In this respect, 4C~74.26 is a particularly
interesting object because its \asca\ X-ray spectrum \citep{b98,sem99}
showed features that are hallmarks of Seyfert galaxies but not BLRGs:
a ``warm'' absorber with a column density of $3.5\times 10^{21} {\rm
cm^{-2}}$ and a significant Compton ``reflection'' hump. It was also
the only QSR in the collection of \citet{sem99} with a detectable
Fe~K$\alpha$ line.

In \S2 we describe the observations and basic data analysis while in
\S3 we present the results of fitting models to the spectra. In \S4 we
discuss our results and we present our final conclusions.
Throughout this paper we adopt a Hubble constant of $H_0=50~{\rm
km~s^{-1}~Mpc^{-1}}$ and a deceleration parameter of $q_0=0.5$.

\section{Observations, Data Screening, and Timing Analysis} 

\asca\ observations of PKS~2349--01 and 3C~323.1 were carried out in
1999 December and 2000 February, respectively. Table~1 reports the
observation log.  The SIS detectors were operated in {\tt 1-CCD FAINT}
mode half of the time and in {\tt 1-CCD BRIGHT} mode for the remaining
portion, while the GIS detectors were operated in {\tt PH} mode. The
data were screened within {\tt FTOOLS/XSELECT v.4.2}
\citep*{bgp94,i94} following the procedures described in \citet{eh98}
and \citet{ehl96}.  The final net exposure times are included in
Table~1.  After screening, we extracted spectra and light curves for
each of the $\asca$ detectors. The \verb+FAINT+ mode data were
converted into \verb+BRIGHT2+ mode and combined to the \verb+BRIGHT+
mode data, to take advantage of the full exposure time and maximize
the signal-to-noise ratio ($S/N$).

4C~74.26 was observed in 1999 May (Table 1) by the \sax\ narrow field
instruments: the Low- and Medium-Energy Concentrator Spectrometers,
and the Phoswich Detector System (hereafter LECS, MECS, and PDS,
respectively). The data from two of the three MECS units were merged
to increase the $S/N$. The exposure times resulting from the
application of standard screening criteria to account for passage
through the South-Atlantic Anomaly, Earth occultation,
etc. \citep{fgg99}, are listed in Table~1. The source was detected in
the PDS band up to energies of order 100~keV, with a count rate of
0.52 $\pm$ 0.04 c~s$^{-1}$ (Table 1).  No bright sources were found in
the LECS and MECS fields of view, thus enhancing our confidence that
the source detected by the PDS is indeed the target quasar.

Figure 1 shows the \asca\ and \sax\ light curves of the three
targets. For PKS~2349--01 and 3C~323.1, we show the 2--10~keV light
curves from the two \asca\ GIS detectors combined, binned at 30-minute
intervals. For 4C~74.26, we show the 2--10~keV light curve from the
MECS instrument, binned at 30-minute intervals, as well as the
15--125~keV light curve from the PDS instrument, binned at 2-hour
intervals.  No systematic variability is evident in any of the light
curves save for a brief, 25\% excursion in the case of
PKS~2349--01. To search for short-time scale fluctuations we applied
the excess variance test \citep{n97a,turner99} to the light curves
binned at both 256-second and 1800-second intervals.  The excess
variance measured from the 256-second light curves is
$(5.4\pm0.1)\times 10^{-2}~{\rm s}^{-1}$, $(5.5\pm0.3)\times
10^{-2}~{\rm s}^{-1}$, and $(4.2\pm0.1)\times 10^{-2}~{\rm s}^{-1}$
for PKS~2349--01, 3C~323.1, and 4C~74.26, respectively. These values
are higher than those of Seyfert galaxies of comparable luminosity
\citep[see, for example][]{leighly99,turner99} but they are similar to
the values found for narrow-line Seyfert~1 galaxies of comparable
luminosity.  The corresponding values of the excess variance for the
1800-second light curves are $(5.8\pm0.1)\times 10^{-2}~{\rm s}^{-1}$,
$(3.7\pm0.1)\times 10^{-2}~{\rm s}^{-1}$, and $(1.8\pm0.1)\times
10^{-2}~{\rm s}^{-1}$ for PKS~2349--01, 3C~323.1, and 4C~74.26,
respectively. The observed fluctuations are unlikely to be related to
emission from the jets of these objects because the radio morphology
suggests that the the jets are oriented at large angles to the line of
sight \citep{a85,k94,r88}. This conclusion is also supported by the
high-energy light curve of 4C~74.26, which is rather steady over the
course of the observation, in marked contrast to the light curves of
blazars.

\section{Model Fits to the Spectra}

Models were fitted to the \asca\ and \sax\ spectra using XSPEC
v.11.0.1 \citep{a96}, in the energy ranges where the calibration is
well established and the background is low (see Table~1). The spectra
were rebinned such that each new bin contained at least 20 counts in
order to validate the use of the $\chi^{2}$ statistic. We used the
most current versions of the response matrices and effective area
curves (September 1, 1997 files for the \sax\ instruments; February 3,
1995 response matrices for the \asca\ GIS, and all other \asca\
calibration files generated specifically for our observations).

To check the self-consistency of the spectra obtained by different
\asca\ detectors, we first fitted the individual SIS and GIS spectra.
We then fitted the same model to the spectra from all detectors
simultaneously. To take account of any uncalibrated differences
between the effective areas of the different instruments, we allowed
the normalization of the corresponding spectra to vary freely.  An
analogous procedure was adopted for the spectra from the \sax\
detectors, as we discuss further below. As with the \asca\ spectra,
the normalization constants of spectra from different instruments were
also allowed to vary, but the PDS normalization was constrained to be
between 0.77 and 0.83 of the MECS normalization. The best-fitting
value for the PDS/MECS ratio was found to be $0.83^{+0}_{-0.06}$. The
best-fitting LECS/MECS ratio, which was left free to vary, was
$0.70\pm$0.02.

The goals of the spectral analysis are to determine the model that
best describes the continuum, including possible excess absorption at
low energies and the Compton hump at $E\simgt 10$~keV, and check for the
presence of the \feka\ emission line. To determine the best-fitting
continuum models, we adopted the following procedure.  At first, the
spectra were fitted by excluding the rest energy range 5--7~keV, where
the \feka\ line contributes most. Once the best-fitting continuum model
was found, the 5--7~keV range was added back to the spectra and a
Gaussian line was added to the model. This procedure ensures the
continuum and the emission line remain unaffected by one another in
the fitting process, an especially important concern if the line is
broad \citep[e.g.,][]{n97b}.

While we investigated the presence of excess absorption at low
energies in the three targets, a detailed analysis was possible only
in the case of 4C~74.26, as the LECS sensitivity extends to lower
energies than the SIS and the spectral calibration is more
certain. For each \asca\ detector we only use data in the energy range
given in Table~1 in order to bypass calibration problems associated
with the SIS efficiency loss at low energies \citep{yaqoob00}.
Excluding the lowest-energy bins in the SIS spectra results in larger
uncertainties in the determination of the absorbing column, however
the column densities turn out to be high enough in both \asca\ targets
that they can be determined fairly well even by the GIS detectors.
The presence of a Compton ``reflection'' hump can be determined with
confidence only in the \sax\ data because of the higher-energy
sensitivity of the PDS compared to the SIS and GIS detectors.

\subsection{The Continuum}

The best-fitting model for the \asca\ spectra of PKS~2349--01 and
3C~323.1 is an absorbed power law with column density in excess of the
Galactic value plus a Gaussian emission line.  The best-fitting
models, parameters, and corresponding error bars (at 90\% confidence
for two parameters of interest) are summarized in Table~2, while the
spectra are shown in Figure~\ref{figspec} with models superposed. All
the models include an absorption component, fixed to the Galactic
column density, and a second absorption component (due to neutral gas)
at the source's redshift with the column density left free to vary in
the fit. The implicit assumption of this procedure is that any excess
absorption is at the source's rest-frame.  The simple power-law model
also provides a good description of the LECS+MECS ($E<10$~keV)
spectrum of 4C~74.26 but it leaves large residuals at high energies,
in the portion of the spectrum sampled by the PDS, as shown in
Figure~\ref{figspec}c.

It is noteworthy that the absorbing column detected in the spectrum of
4C~74.26 is rather high, namely $3\times 10^{21}~{\rm cm}^{-2}$, a few
times higher than what is measured for the other two objects. Analysis
of the {\it ASCA} spectrum of this object by \citet{b98} suggested the
presence of an intrinsic, ionized absorber (``warm'' absorber). That
result, however, was not confirmed in the independent analysis of the
same data by \citet{sem99}. To investigate this question further we
fitted the LECS spectrum with two different models representing an ionized
absorber: (a) a single power law with Galactic and intrinsic
absorption, plus a redshifted absorption edge, and (b) an ionized
absorber model according to \citet[][implemented by the XSPEC routine
{\tt absori}]{d92}. In neither case did we find a significant
improvement of the fit: the total value of $\chi^2$ decreases because
of the additional free parameters of these more sophisticated
absorption models, but the improvement is significant only at the
50--65\% level (according to the F-test). Our conclusion, therefore,
is that although a warm absorber provides an acceptable fit, it is not
required by the data.

The high-energy residuals in the simple power-law fit to the spectrum
4C~74.26 indicate the presence of an additional continuum component
emerging at energies \simgt 10~keV.  The following two possibilities
were considered in detail:

\begin{enumerate}

\item
The high-energy excess flux is due to Compton ``reflection'' of the
primary continuum by neutral, dense matter in the immediate vicinity of
the X-ray source \citep{lw88, gr88}. Thus, we fitted the \sax\
spectrum with a Compton ``reflection'' model \citep[{\tt pexrav}
within XSPEC][]{mz95}. This model has several additional free
parameters, which include the inclination angle, $i$, of the
reprocessing slab relative to the line of sight, the solid angle,
$\Omega$, that the slab subtends to the primary X-ray source, and the
upper cut-off energy of the primary power-law spectrum. 
In the particular case of 4C~74.26 we can use the radio properties to
constrain the inclination angle of the radio jet (hence the
inclination angle of the disk). The brightness contrast between the
small-scale jet and counter-jet (observed by VLBI), if ascribed to
relativistic beaming, yields an upper limit to the jet inclination
angle of $i\leq 49^{\circ}$ \citep{pbrw92}. Requiring that the size of
the large-scale radio source does not exceed the size of giant radio
sources observed in radio galaxies can yield a lower limit on the
inclination. Such giant radio source sizes typically range up to
2.0--2.5~Mpc \citep[e.g.,][]{palma00,lara01}. Assuming that the radio
source in 4C~74.26 is not exceptionally large and adopting 2~Mpc as an
upper limit to its true size, we obtain $i\geq 37^{\circ}$. Thus we
can express the inclination angle as $i=43\pm 6$~degrees.
\footnote{The constraints on the inclination angle of
the jet derive from ratios of fluxes or lengths; thus they do
not depend on $H_0$.}
In Table~2 we give the best fitting model parameters for the case where the
inclination angle of the axis of the reprocessing slab was held fixed
to the value inferred from the radio morphology.  In
Figure~\ref{figrefl} we show how the inferred reprocessor solid angle
depends on the value of the inclination by plotting the 68, 90, and
99\% confidence contours in the $\cos i - \Omega/2\pi$ plane. For
inclination angles between $37^{\circ}$ and $49^{\circ}$, which
represent the range inferred from the radio properties, the solid
angle is consistent with $2\pi$. The fit with the Compton
``reflection'' model is an improvement over the single power law at the
98.7\% confidence level, according to the F-test.

\item
The excess flux could indicate a contribution from the base of the
jet. This was modeled by adding a second power law to the continuum,
which did not yield a significantly different description, however,
leading us to disfavor this option. More specifically, the spectral
indices of the two power-law components converge to values that are
indistinguishable from each other and the total $\chi^2$ value is
higher than the Compton ``reflection'' model, which has more free
parameters. Moreover, emission from a jet is disfavored by the radio
properties of this object (mainly the morphology), which suggest that
the jet is oriented at a large angle to the line of sight
($i>37^{\circ}$).

\end{enumerate}

Although the \asca\ bandpass is not well-suited to the detection of
the Compton ``reflection'' continuum, we nevertheless searched for it
in the spectra of PKS~2349--01 and 3C~323.1. The main reason for
embarking on this exercise is that the adopted continuum model could
have an influence on the measured properties of the \feka\ lines,
which we study in the next section (see the discussion in the context of
4C~74.26 in \S3.3). In the case of 3C~323.1 we find no evidence for a
Compton ``reflection'' continuum. In fact, a Compton ``reflection''
model converges to a model with negligible ``reflection,'' which is in
effect the original simple power-law model. In the case of
PKS~2349--01 the Compton ``reflection'' model results in a slightly
lower $\chi^2$ value than the simple power-law model, but this
apparent improvement is not significant (the F-test gives a 66\%
probability of improvement by chance). But the two continuum models
yield significantly different properties for the \feka\ line, as we
describe below. This effect has been known for some time, and has been
discussed by a number of authors, including most recently
\citet{zg01}.

\subsection{The Fe K$\alpha$ Emission Line}

To search for the presence of an Fe~K$\alpha$ emission line, which is
a ubiquitous feature of the X-ray spectra of Seyfert galaxies, we
modeled its profile with a Gaussian. We fitted the continuum as
described above and fixed the continuum parameters before adding the
Gaussian line profile to the model. To evaluate whether the line is
unambiguously detected and resolved we searched the 2-dimensional
parameter space defined by the line photon flux and the energy
dispersion of the Gaussian profile. The results of this search are
shown in Figure~\ref{figlines} in the form of confidence contours in
the line EW $vs$ FWHM plane (the EW and FWHM are directly proportional
to the line photon flux and Gaussian energy dispersion, respectively).
These results are also summarized in Table~3.

In the case of 4C~74.26 the line is clearly detected with a rest EW of
$170^{+180}_{-100}$~eV. The energy of the Fe~K$\alpha$ emission line
encompasses the range 6.4 to 6.9~keV within 90\% confidence.  Thus, we
confirm the detection of the line in 4C~74.26 with \asca, as reported
by \citet{b98} and \citet{sem99}. In the case of 3C~323.1 the line is
only marginally detected, i.e., its EW is non-zero at 90\% confidence
but consistent with zero at 99\% confidence (for an assumed energy of
6.4~keV). The best-fitting FWHM is about $5\times 10^4~{\rm
km~s}^{-1}$, which is comparable with what is found in Seyfert
galaxies by \citet{n97b}.

In the case of PKS~2349--01, the measured properties of the \feka\
line depend sensitively on the continuum model. If we adopt a simple
power-law model, the line is unambiguously detected at the 99\%
confidence level with a rest EW of $230\pm 120$~eV and a FWHM of
$6^{+6}_{-3}\times 10^4~{\rm km~s}^{-1}$. If, however, we adopt a
model consisting of a power-law plus Compton ``reflection,'' fixing
the solid angle of the reprocessor to $\Omega=2\pi$, the line is about
a factor of 2 weaker (see Table~3). If we fix the solid angle of the
reprocessor to $\Omega=4\pi$, its maximum physically reasonable value,
the line is only marginally detected (at the 68\% confidence
level). The upper limit to its rest EW is 200~eV, assuming a FWHM of
$5\times 10^4~{\rm km~s}^{-1}$. This ambiguity cannot be resolved with
the data at hand. In order to measure the properties of the
Fe~K$\alpha$ lines with confidence, we need either a 2--10~keV
spectrum of much higher $S/N$ or a spectrum of comparable $S/N$
extending to at least 30~keV, which can yield meaningful constraints
on the properties of the Compton ``reflection'' component.  Finally,
we also note that the observed EWs of the Fe~K$\alpha$ lines and the
strength of the Compton ``reflection'' components are consistent with
each other, albeit within the large uncertainties, as one would expect
in the context of reprocessing models such as those of
\citet{gf91}. Such models attribute the origin of both the
Fe~K$\alpha$ line and the Compton ``reflection'' component to the same
reprocessing medium and require that their strengths are related.

\subsection{Comparison With Previous Observations of 4C~74.26 by ASCA}

Our preferred model for 4C~74.26 consists of a power-law continuum
modified by Compton ``reflection'' at high energies and by absorption
at low energies. The line energy obtained from the model could not be
tightly contrained, falling between 6.4 and 6.9~keV at a 90\%
confidence level. Within uncertainties, our results are largely
consistent with the results of \citet{b98}, who considered our
preferred model among others.  However, there is a systematic
discrepancy between the properties of the Fe~K$\alpha$ line, which we
attribute to the continuum fit used by \citet{b98}. More specifically
we find that the line is several times broader and about 70\% stronger
than what \citet{b98} found. Comparing the parameters describing the
continuum, we see that the Compton ``reflection'' component in the
{\it ASCA} spectrum of Brinkmann et al. is poorly constrained and has
an extremely high value that implies the reprocessor subtends an
area larger than 3 times the area of the sky.  Although this could be
interpreted as an indication that the primary X-ray source does not
emit isotropically, it is also possible that the strength of the
Compton ``reflection'' component was overestimated.

Such poor constraints on the Compton ``reflection'' component are not
surprising, since the \asca\ sensitivity drops dramatically at
$E>8$~keV, making it practically impossible to measure this
component. Thus, the continuum around the Fe~K$\alpha$ line is
elevated, taking away the broad wings of the line and reducing its
overall strength. In contrast, in our \sax\ spectrum we are able to
determine the shape of the high-energy continuum unambiguously with
the help of the PDS. We find a considerably weaker Compton
``reflection'' strength than \citet{b98} and a correspondingly broader
and stronger Fe~K$\alpha$ line.

In view of the strength of the Compton ``reflection'' component
measured from the \sax\ spectrum, we re-analyzed the the \asca\
spectra to check for consistency. We extracted and fitted the spectra
with the same methodology described in \S2 and \S3, keeping the
Compton ``reflection'' component fixed to a strength corresponding to
$\Omega=2\pi$, as measured from the \sax\ spectrum. We measured the EW
of the Fe~K$\alpha$ line for fixed rest energies of 6.4 and
6.7~keV. We found that the measurements from the \asca\ spectra are
consistent with those from the \sax\ spectrum.  In particular, we
measure a spectral index of $1.87_{-0.07}^{+0.08}$, while the
Fe~K$\alpha$ rest EW is poorly constrained and lies in
the range 130--700 eV, as listed in Table~3.

\section{Summary and Discussion}

We have presented new \sax\ and \asca\ observations of three luminous
($L_{2-10~{\rm keV}}\sim 10^{45}~\lum$) radio-loud AGN, as part of our
ongoing systematic study of the X-ray properties of such objects. We
find that in all three cases the main underlying continuum can be
described as a power law with photon index $\Gamma \approx 1.85$, plus
excess absorption. In two sources (PKS~2349--01 and 4C~74.26) the
\feka\ emission line is detected with rest-frame ${\rm EW} \sim$
200~eV but with large uncertainties, of a factor of 2; in 4C~74.26 the
line is unresolved while in PKS~2349--01 the line properties depend on
the model adopted for the continuum.  In 4C~74.26, a strong ($\Omega
\approx 2\pi$) Compton ``reflection'' component is also detected with
the PDS instrument on \sax. In PKS~2349--01, on the other hand, the
data allow the presence of a Compton ``reflection'' component but
cannot provide meaningful constraints on its strength. This is
unfortunate because the strength of this component affects the
inferred properties of the \feka\ line dramatically, as we explained in
detail in \S3.2.

The average photon index for the three sources studied here is
$\langle\Gamma\rangle=1.86$ with a small dispersion of
$\sigma_{_{\Gamma}}=0.01$. The average slope of the X-ray continuum is
consistent with the average photon index previously measured with
\asca\ for a small sample of 7 QSRs of comparable luminosity, which
had $\langle\Gamma\rangle=1.80$ and $\sigma_{_{\Gamma}}=0.15$
\citep{sem99}. This is also consistent with the average slope for a
sample of radio-quiet quasars from \asca\ measurements,
$\langle\Gamma\rangle=1.86$ and dispersion $\sigma_{_{\Gamma}}=0.20$
\citep[see Table~5 of][]{sem99}. The similarity of slopes for the
radio-loud and radio-quiet sources confirms that no contribution from
the beamed emission of an unresolved jet is present in the radio-loud
QSRs of our study (as expected based on their radio morphology, see
above). It also suggests a similar continuum emission process in the
two classes of AGN. This result will need to be confirmed using larger
samples of both radio-loud and radio-quiet sources. Moreover, a wider
spectral coverage will also be needed in order to constrain the
strength of the Compton ``reflection'' hump and disentangle its effects
from the slope of the intrinsic continuum.

Another result of this paper is the excess absorption found in
all three objects. In 4C~74.26, where the observed spectra extend to
low energies, we have searched for the signature of a warm absorber,
but have not found conclusive evidence for it, nor were we able to
rule it out. There are two possible origins for the excess
absorption. The first is local excess absorption in the Galaxy. As
4C~74.26 lies at relatively low latitudes ($b=19^{\circ}\!\!.5$), it
is conceivable that the extra absorption is associated with a
Galactic molecular cloud. Alternatively, the absorber could be
intrinsic to the source. In PKS~2349--01 and 3C~323.1, the absorber is
most likely intrinsic to the source because of their high Galactic
latitudes. This is in agreement with earlier findings \citep{sem99}
that excess neutral absorption is common in radio-loud AGN and the
absorbing column does not appear to be correlated with the X-ray
luminosity of the source \citep[these three objects fall in the
general area occupied by radio-loud AGNs in the $N_{\rm H}~vs~L_{\rm
X}$ plot shown in Figure~6 of][]{sem99}.

We have also detected an \feka\ emission line in two (possibly three)
sources of our sample.  In order to compare the strengths of these
lines with what is found in other radio-loud and radio-quiet AGNs we
plotted the rest EWs against the 2--10~keV luminosity in
Figure~\ref{figewlx}. This is an update of the figure presented in
\citet{sem99} and \cite{esm00}. The data points represent radio-loud
AGNs \cite[][and this paper]{sem99,esm00,e02}. The shaded area
represents the ``X-ray Baldwin effect'' for radio-quiet Seyferts and
QSOs and its 90\% dispersion \citep{n97c}.  While at $L_{\rm
X}<5\times 10^{44}~\lum$ radio-loud AGNs have systematically lower
\feka\ EWs than radio-quiet AGNs of comparable luminosity, the
situation at higher luminosities is quite unclear because of the large
uncertainties in the measured EWs. There are two possible ways of
viewing Figure~\ref{figewlx}: either (a) radio-loud AGNs follow their
own ``X-ray Baldwin effect,'' which is systematically offset from that
of radio-quiet AGNs, or (b) radio-loud AGNs do not show a Baldwin
effect at all. Unfortunately, the large uncertainties in the measured
EWs, especially at high luminosities, do not allow us to distinguish
between these two possibilities. Better measurements of the \feka\ EWs
of luminous radio-loud AGNs ($L_{\rm X}> 10^{45}~\lum$) are sorely
needed.

The X-ray Baldwin effect for radio-quiet AGN was interpreted by
\citet{n97c} as the result of a progressive ionization of the inner
accretion disk as the X-ray luminosity increases. We speculate that
the lack of a similar trend for radio-loud AGN indicates that the
ionization structure of the disk of radio-loud AGNs does not change
significantly as the X-ray luminosity increases. This could happen,
for example, in a scenario where the dense, optically thick matter in
the disk is photoionized by external illumination with a hard spectrum
of low ionization parameter. If the ionization parameter in the \feka\
emitting region is extremely low, changes in the ionizing luminosity
even by an order of magnitude are accompanied by a proportional
increase in the line luminosity. This would be the case if the central
accretion disk is an ADAF (or similar structure) which has a low
radiative efficiency and produces an X-ray luminosity that is
disproportionately low compared to the accretion rate and a relatively
hard spectrum. Ionization of the geometrically thin disk exterior to
the ADAF by hard X-rays from the ADAF itself would thus produce the
desired effect.

The above picture is supported by a number of other observational
results, namely the systematically lower \feka\ EWs and FWHMs and
weaker Compton ``reflection'' humps of radio-loud AGNs compared to
their radio-quiet counterparts \citep[see][]{w98,esm00,g02}. Both of
these differences can be understood in the context of a scenario in
which the inner accretion disk is an ADAF rather than a geometrically
thin and optically thick flow. The difference in \feka\ EWs and
Compton ``reflection'' strengths is then a geometrical effect
resulting from the small solid angle subtended by the outer disk to
the primary X-ray source \citep*{ch89,zls99}. The difference in \feka\
FWHM follows simply from Kepler's law.

To make further progress on this problem, more and better hard X-ray
spectra of radio-loud AGN are needed, especially at high
luminosity. The ideal instrument for this job is \xmm. These spectra
will serve to test the X-ray Baldwin effect for this class of objects.
An independent test of the above hypothesis will be afforded by the
profiles of the \feka\ lines, which would be expected to be
significantly different in width and asymmetry from those of Seyfert
galaxies.

\acknowledgements

We thank the anonymous referee for very helpful comments.  CAH was
funded by a fellowship from the Schreyer Honors College of the
Pennsylvania State University and by NASA grants NAG5-9133 and
NAG5-27017. This work was also partially supported by NASA grants
NAG5-8369 and NAG5-10817.  We have made use of the NASA/IPAC
Extragalactic Database (NED) which is operated by the Jet Propulsion
Laboratory, California Institute of Technology, under contract with
the National Aeronautics and Space Administration.


\clearpage


\singlespace
\begin{deluxetable}{ccrlccc}
\tablenum{1}
\tablewidth{5.5in}
\tablecolumns{7}
\tablecaption{Target Properties and Observation Details}
\tablehead{
\colhead{}              &
\colhead{Galactic}      &
\colhead{}              &
\colhead{Energy}        &
\colhead{Expos.}      & 
\colhead{Count}         &
\colhead{Start Date}    \\
\colhead{}              &
\colhead{$N_{\rm H}$}   &
\colhead{}              &
\colhead{Band}          &
\colhead{Time}          &
\colhead{Rate}          &
\colhead{and Time}      \\
\colhead{$z$}           &
\colhead{(${\rm cm}^{-2}$)}   &
\colhead{Instrument}    &
\colhead{(keV)}         &
\colhead{(ks)}          &
\colhead{(s$^{-1}$)}    &     
\colhead{(UT)}          
}
\startdata
\cutinhead{PKS 2349--01}
 0.174 & $3.26 \times10^{20}$ & \asca\ GIS2 & 0.95--10  & \phantom{0}71.47  & 0.149$\pm$0.002  & 1999/12/24 \\  	   
       &                      &        GIS3 & 0.95--10  & \phantom{0}71.45  & 0.188$\pm$0.002  & 05:24 \\ 
       &                      &        SIS0 & 0.85--8   & \phantom{0}62.36  & 0.272$\pm$0.002  & \\   
       &                      &        SIS1 & 1.35--8   & \phantom{0}62.30  & 0.220$\pm$0.002  & \\
\cutinhead{3C 323.1}
 0.264 & $2.32 \times10^{20}$ & \asca\ GIS2 & 0.95--10  & \phantom{0}41.84  & 0.125$\pm$0.002  & 2000/02/08 \\
       &                      &        GIS3 & 0.95--10  & \phantom{0}41.86  & 0.152$\pm$0.002  & 00:35 \\
       &                      &        SIS0 & 0.85--8   & \phantom{0}37.37  & 0.222$\pm$0.003  & \\
       &                      &        SIS1 & 1.35--8   & \phantom{0}37.46  & 0.184$\pm$0.003  & \\
\cutinhead{4C 74.16}
 0.104 & $1.19 \times10^{21}$ & $SAX$  LECS & 0.40--4   & \phantom{0}44.50  & 0.161$\pm$0.002  & 1999/05/17 \\
       &                      &        MECS & 1.65--10.5 &          100.28  & 0.308$\pm$0.002  & 02:53 \\
       &                      &         PDS & 15--125    &          100.00  &  0.52$\pm$0.04  & \\

\enddata
\end{deluxetable}

\begin{deluxetable}{lccccc}
\tablenum{2}
\tablewidth{5.5in}
\tablecolumns{6}
\tablecaption{Best-Fitting Continuum Models and Parameters}
\tablehead{
\colhead{}						      &
\multicolumn{2}{c}{Observed Flux\tablenotemark{b}}            &
\multicolumn{2}{c}{Luminosity\tablenotemark{b}}     \\
\colhead{}                                                    &
\multicolumn{2}{c}{($10^{-12}~{\rm  erg~s^{-1}~cm^{-2}}$)}    &
\multicolumn{2}{c}{($10^{45}~{\rm erg~s^{-1}}$)}              \\
\noalign{\vskip -8pt}					   
\colhead{Best-Fitting Models}                                 &
\multicolumn{2}{c}{\hrulefill}                                &
\multicolumn{2}{c}{\hrulefill}                                &
\colhead{$\chi^2_{\rm r}$}                                    \\
\colhead{and Parameters\tablenotemark{a}}                     &
\colhead{0.5--2 keV}				              &
\colhead{2--10 keV}					      &
\colhead{0.5--2 keV} 					      &
\colhead{2--10 keV}    	                                      &
\colhead{(d.o.f.)}                                            
}
\startdata
\cutinhead{PKS~2349--01}
Absorbed power law                                                         & 3.3 & 6.8  & 0.47 & 0.97 & 1.05 \\  
$N_{\rm H}^{\rm exc.} = (9\pm 5) \times 10^{20}{\rm cm}^{-2}$              & & & & & (250) \\
$\Gamma$ = 1.87$\pm{0.04}$                                                 & & & & &  \\
\cutinhead{3C~323.1}							 
Absorbed power law                                                         & 3.2 & 5.5 & 1.09 & 1.87 & 1.12 \\
$N_{\rm H}^{\rm exc.} = (6\pm 5)\times 10^{20}{\rm cm}^{-2}$               & & & & & (278) \\
$\Gamma$ =  1.86 $^{+0.02}_{-0.04}$                                        & & & & & \\
\cutinhead{4C~74.26}							 
Absorbed power law                                                         & 15.1 & 14.1 & 0.74 & 0.69 & 1.07 \\
plus Compton reflection                                                    & & & & & (105) \\
$N_{\rm H}^{\rm exc.} = (3.0\pm 0.3)\times10^{21}{\rm cm}^{-2}$            & & & & & \\
$\Gamma$ = 1.85$\pm^{+0.05}_{-0.04}$                                       & & & & & \\
\multicolumn{6}{l}{Reprocessor Solid Angle, $\Omega/2\pi$ = 1.2$^{+0.6}_{-0.5}$} \\
\multicolumn{6}{l}{Reprocessor Inclination Angle\tablenotemark{c}, ${\it i}$= 43$^\circ$} \\
\multicolumn{6}{l}{Power-Law Folding Energy $> 140$~keV} \\
\tablenotetext{a\;}{All error bars correspond to 90$\%$ confidence limits. The quoted absorbing column is in excess of the
Galactic column and is assumed to be due to neutral gas at the redshift of the source.} 
\tablenotetext{b\;}{The observed flux is not corrected for absorption, while the luminosity is corrected for absorption. All bandapasses refer to 
the observer's frame.}
\tablenotetext{c\;}{The inclination angle of the disk was inferred from the radio morphology and held fixed for the purpose of 
estimating the error bars given here. Its effect on the measured value of $\Omega/2\pi$ is explored separately in Figure~\ref{figrefl}.}

\enddata
\end{deluxetable}


\singlespace 
\begin{deluxetable}{lcccl}
\tablenum{3}
\tablewidth{5.8in}
\tablecolumns{5}
\tablecaption{Gaussian Model Parameters for Fe K$\alpha$ Emission Lines\tablenotemark{a}}
\tablehead{
\colhead{}                & 
\colhead{Rest Energy}     & 
\colhead{FWHM} 	& 
\colhead{Rest EW} &
\colhead{} \\ 
\colhead{Telescope}                & 
\colhead{(keV)}     & 
\colhead{($10^4$~km s$^{-1}$)}    & 
\colhead{(eV)}  & 
\colhead{Continuum Model}  
}  
\startdata
\cutinhead{PKS~2349--01 }
\asca\       &  $6.2^{+0.3}_{-0.2}$ &  $6^{+6}_{-3}$ & $230\pm 120$ & simple power law\\ 
\asca\       &  $6.2^{+0.3}_{-0.2}$ &  $4\; (<15)$   & $90^{+90}_{-70}$ & power law + ``reflection'' ($\Omega=2\pi$) \\
\asca\       &  $6.2^{+0.3}_{-0.2}$ &  unconstr.     & $<200$ \,\tablenotemark{b} & power law + ``reflection'' ($\Omega=4\pi$) \\
\cutinhead{3C~323.1}
\asca\   &  6.4 (fixed)  & $<13$\,\tablenotemark{c} &  $90^{+90}_{-80}$\,\tablenotemark{c}  & simple power law  \\
\cutinhead{4C~74.26}
\sax\  &  $6.7^{+0.2}_{-0.4}$ &  $5\; (<15)$  &  $170^{+180}_{-100}$ & power law + ``reflection'' (Table 1) \\
\asca\,\tablenotemark{d}\ &  6.4 (fixed) & $60^{+120}_{-30}$ & $250^{+400}_{-120}$  & power law + ``reflection''  ($\Omega=2\pi$)   \\
\asca\,\tablenotemark{d, e}\ &  6.7 (fixed) & $80^{+100}_{-80}$ & $290^{+410}_{-140}$  &  power law + ``reflection''  ($\Omega=2\pi$)   \\
\tablenotetext{a\;}{All error bars and upper limits correspond to the 90$\%$ confidence intervals, unless otherwise noted.}
\tablenotetext{b\;}{The upper limit to the rest EW in the latter case corresponds to a FWHM of 
$5\times 10^4~{\rm km~s}^{-1}$.}
\tablenotetext{c\;}{In the case of 3C~323.1 a line is detected only at 90\% confidence and only when we fix the rest energy
at 6.4~keV. The FWHM is unconstrained at 90\% confidence; the quoted limit corresponds to the 68\% confidence interval.
The quoted value of the EW is based on the assumption that ${\rm FWHM=5\times 10^4~km~s^{-1}}$.}
\tablenotetext{d\;}{From the re-analysis of the data already published by \citet{b98} and \citet{sem99}. See \S3.3 for details.}
\tablenotetext{e\;}{The line properties are unconstrained at 90\% confidence. The error bars given here correspond to 68\% confidence.}
\enddata
\end{deluxetable}

\clearpage

\begin{figure}
\plotone{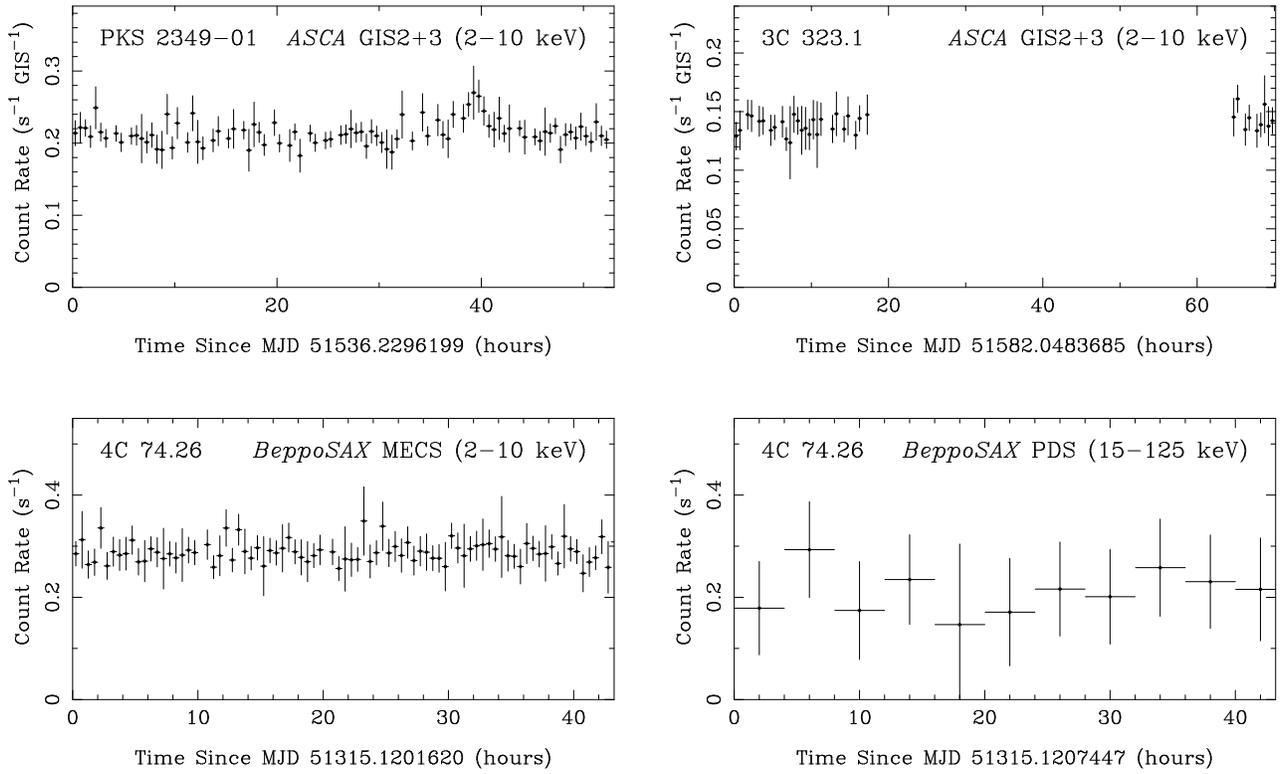}
\caption{\label{figlc} Light curves of the three quasars. The light
curves of PKS~2349--01 and 3C~323.1 show the average 2--10~keV count
rate from the two GIS instruments. In the case of 4C~74.26 we show a
separate light curve from two of the \sax\ instruments: 2--10~keV
(MECS) and 15--125~keV (PDS).}
\end{figure}

\begin{figure}
\plotone{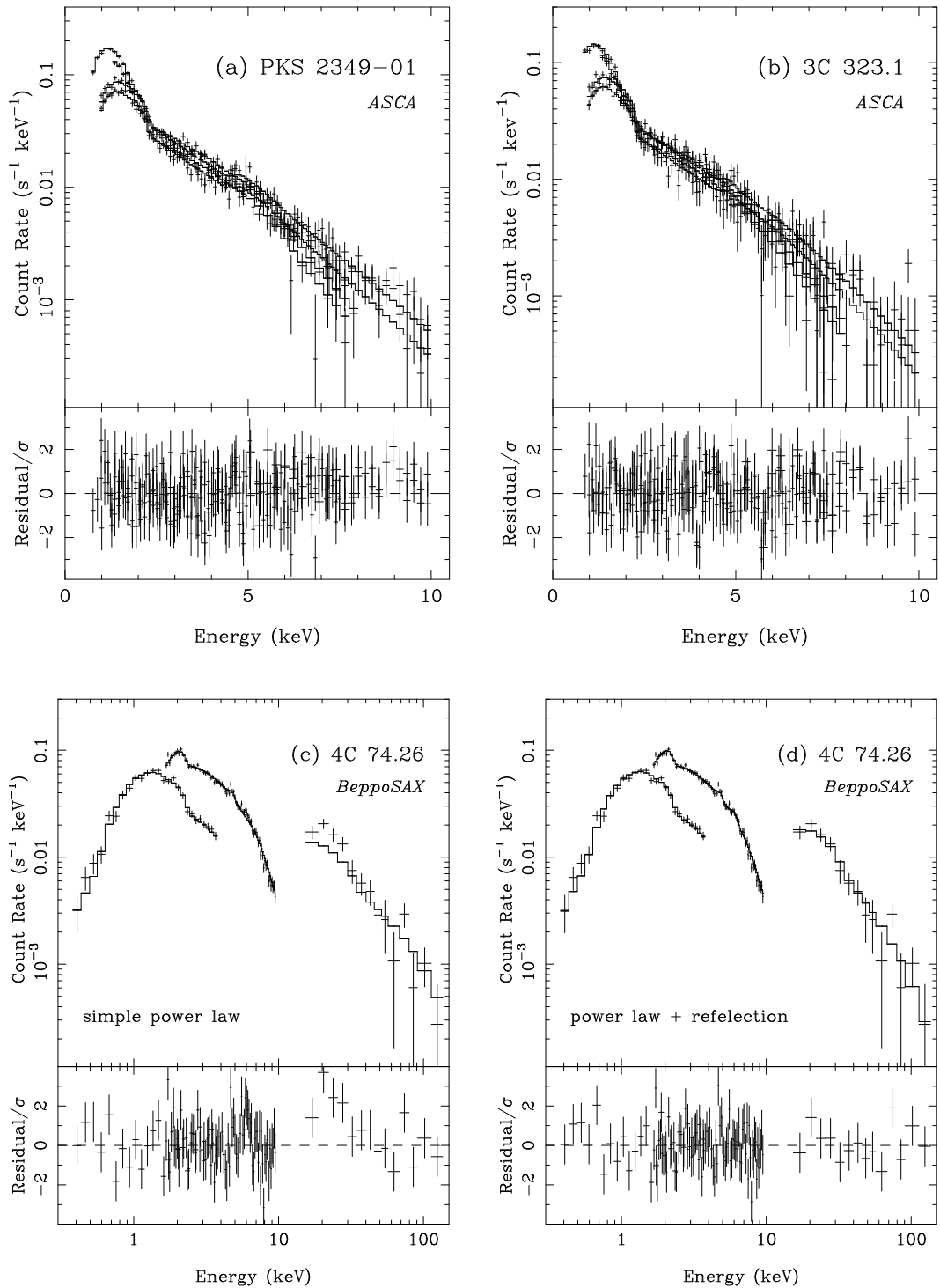}
\caption{\label{figspec} Spectra of the three quasars with models
superposed.  In panels (a) and (b) we show fits to the \asca\ spectra of
PKS~2349--01 and 3C~323.1 with a model consisting of an absorbed power
law plus a Gaussian emission line.  In panel (c) we show a simple
power-law fit to the \sax\ spectrum of 4C~74.26, which leaves large
residuals, especially at high energies. Panel (d) shows the \sax\
spectrum of 4C~74.26 with a model consisting of an absorbed power law
plus Compton ``reflection'' and a Gaussian \feka\ line which yields a
considerably better fit than the simple power-law model.}
\end{figure}

\begin{figure}
\plotone{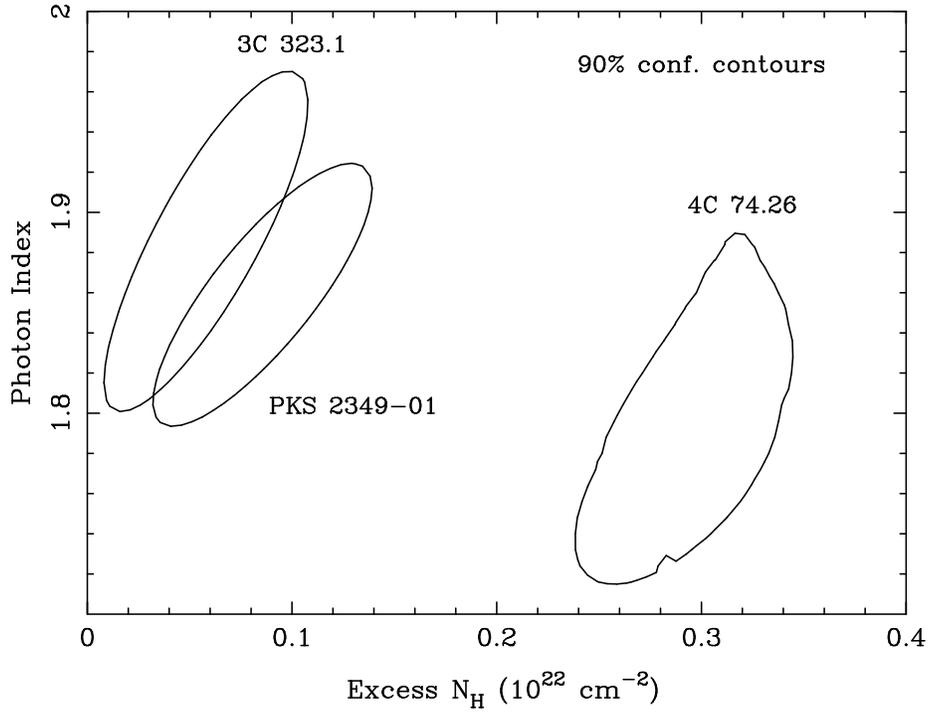}
\caption{\label{figcont} Contours showing the 90\% confidence
intervals in the excess absorption $vs$ photon index parameter
plane. The excess absorbing column is in addition to the Galactic
column listed in Table~1.  The contours for PKS~2349--01 and 3C~323.1
were derived by fitting the spectra from all \asca\ instruments
simultaneously, while the contours for 4C~74.26 were derived by
fitting the \sax\ LECS and MECS spectra simultaneously.}
\end{figure}

\begin{figure}
\plotone{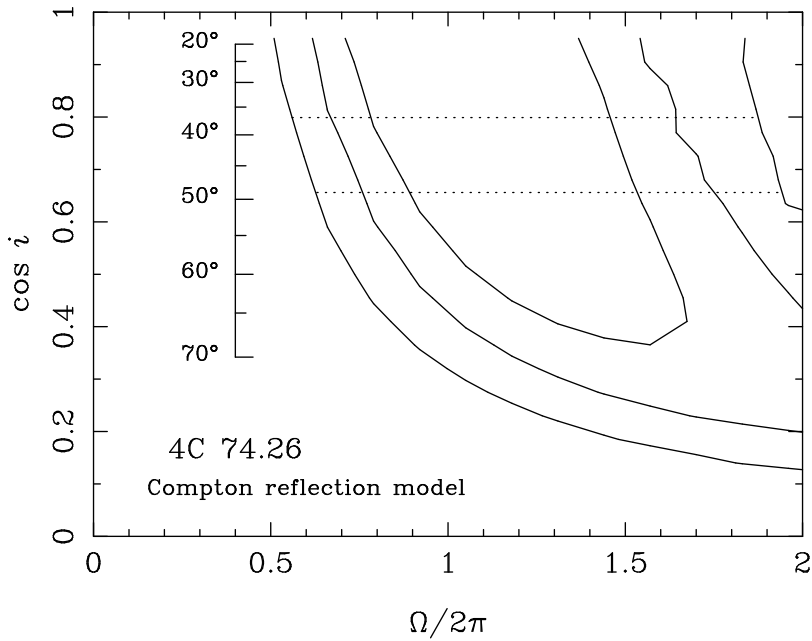}
\caption{\label{figrefl} Contours showing the 68, 90, and 99\%
confidence intervals in the inclination angle $vs$ solid angle
parameter plane for the Compton ``reflection'' model applied to
4C~74.26. The primary power-law index and absorbing column density 
were allowed to vary during the fit but the upper cut-off energy of the 
primary spectrum was held fixed at 140~keV. The horizontal dotted lines show 
the range of inclination angles allowed by the radio properties (see the 
discussion in \S3.1 of the text).}
\end{figure}

\begin{figure}
\plotone{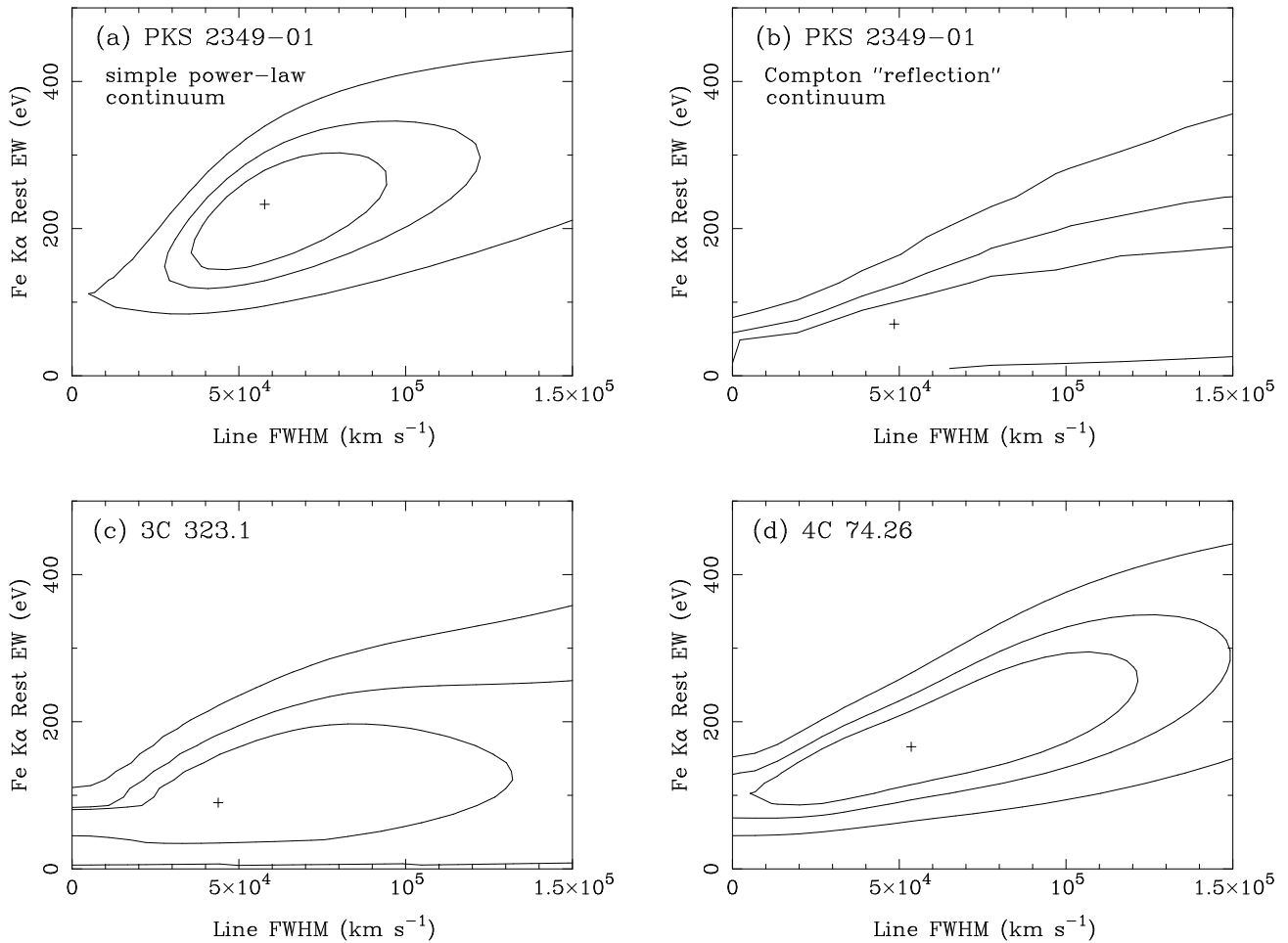}
\caption{\label{figlines} Contours showing the 68, 90, and 99\%
confidence intervals in the rest EW $vs$ FWHM plane for the Gaussian
emission-line model applied to the three quasars. The confidence
intervals were obtained after fixing the parameters describing the
continuum (see \S3.1 of the text for details).  The two panels for
PKS~2349--01 refer to different continuum models: a simple power-law
and a Compton ``reflection'' model. The panel for 3C~323.1 refers to s
simple power-law model, while the panel for 4C~74.26 refers to a
Compton ``reflection'' model. Additional details are given in the text.}
\end{figure}

\begin{figure}
\plotone{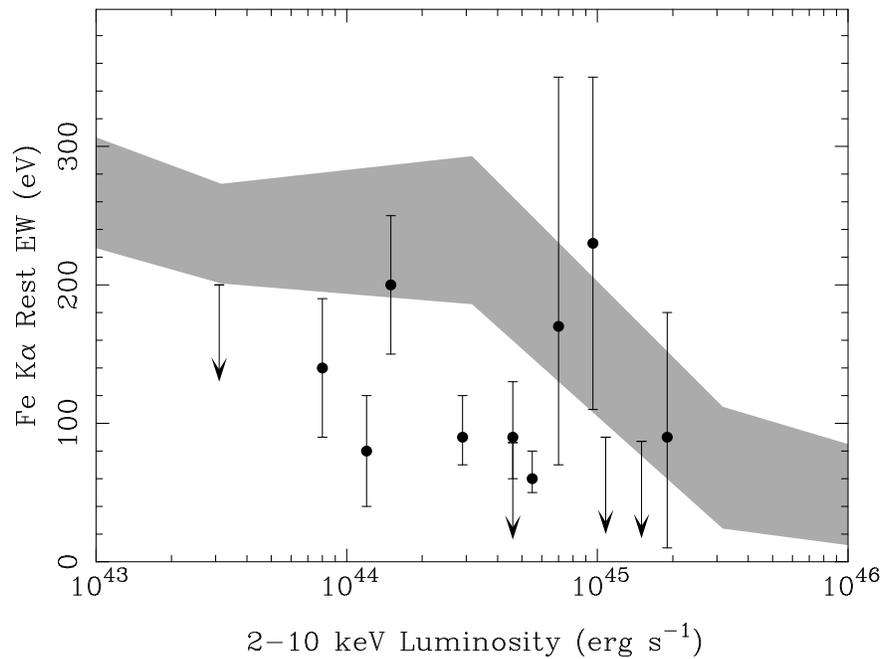}
\caption{\label{figewlx} Comparison of the trends between the
Fe~K$\alpha$ rest EW and 2--10~keV luminosity in radio-quiet and
radio-loud AGNs. The grey band shows the region of this diagram
occupied by the sample of \cite{n97c}, which consists primarily of
radio-quiet AGNs at $L_{\rm X} < 10^{45}~\lum$. The data points
represent radio-loud AGNs from this paper and earlier papers in our
program, as follows: Arp~102B \cite[upper limit at $L_{\rm X} =
3\times 10^{43}~\lum$; from][]{e02}, 3C~445 and 3C~390.3 \cite[${\rm
EW}\approx 140$ and 200 eV; from][]{sem99}, Pictor A, 3C~120, 3C~382,
3C~111 \cite[${\rm EW}\simlt 100$~eV and $10^{44}~\lum<L_{\rm
X}<10^{45}~\lum$; from][]{esm00}, 3C~249.1, 3C~109, 3C~254
\citep[upper limits at $4\times 10^{44}~\lum<L_{\rm X}<2\times
10^{45}~\lum$; from][]{sem99}, 4C~74.26, PKS~2349--01, and 3C~323.1
(large error bars at $L_{\rm X} > 7\times 10^{44}~\lum$; from this
paper). At $L_{\rm X} < 5 \times 10^{44}~\lum$ the radio-loud AGNs
fall systematically below the trend defined by the radio-quiet AGNs,
while at higher luminsoities the situtation is unclear.}
\end{figure}

\end{document}